\documentclass[aps,prl,longbibliography,preprint]{revtex4-1}

\usepackage[utf8]{inputenc}
\usepackage{amsmath, amssymb,graphicx}
\usepackage[cdot,mediumqspace,squaren]{SIunits}
\usepackage{hyperref}
\renewcommand{\vec}{\mathbf} % Vecteur
\newcommand{\e}[1]{\text{e}^{#1}} % exponentielle
\newcommand{\eps}{\varepsilon} % epsilon
\newcommand{\db}[1]{\overline{#1}} % double barre (dyadic)
\graphicspath{{./images/},{./},{../images/}}
\begin{document}

\title{Nonreciprocal and non-Hermitian material response inspired by semiconductor transistors}
\author{Sylvain Lanneb\`{e}re\textsuperscript{1}}
\author{David E. Fernandes\textsuperscript{1}}
\author{Tiago A. Morgado\textsuperscript{1}}
\author{M\'{a}rio G. Silveirinha\textsuperscript{2}}
\email{To whom correspondence should be addressed:
mario.silveirinha@co.it.pt}
 \affiliation{\textsuperscript{1}
Instituto de Telecomunica\c{c}\~{o}es and Department of Electrical Engineering, University of Coimbra, 3030-290 Coimbra, Portugal}
\affiliation{\textsuperscript{2}University of Lisbon -- Instituto
Superior T\'ecnico and
Instituto de Telecomunica\c{c}\~{o}es, Department of Electrical Engineering, 1049-001
Lisboa, Portugal}

\begin{abstract}
Here, inspired by the operation of conventional semiconductor transistors, we introduce a novel class of bulk materials with nonreciprocal
and non-Hermitian electromagnetic response. Our analysis shows that
material nonlinearities combined with a static electric bias may
lead to a \textit{linearized} permittivity tensor that lacks the Hermitian
and transpose symmetries. Remarkably, the material can either dissipate or generate energy, depending on the
relative phase of the electric field components. We introduce a
simple design for an electromagnetic isolator based on an idealized
``MOSFET-metamaterial'' and show that its performance can in
principle surpass conventional Faraday isolators due to the material
gain.
 Furthermore, it is suggested that analogous
material responses may be engineered in natural media in nonequilibrium situations. Our solution
determines an entirely novel paradigm to break the electromagnetic
reciprocity in a bulk nonlinear material using a static electric
bias.
\end{abstract}

\maketitle

%\section{Introduction}
The Lorentz reciprocity law constrains the propagation of
electromagnetic waves in conventional photonic platforms
\cite{pozar_microwave_2011,caloz_electromagnetic_2018}. For
Hermitian systems, the Lorentz reciprocity is rooted in the
linearity of Maxwell's equations and on their invariance under a
time-reversal (TR) operation \cite{silveirinha_time-reversal_2019}.
Generically, a nonreciprocal electromagnetic response requires
either (i) breaking the TR symmetry with a suitable bias, or (ii) using
nonlinear materials, or (iii) exploiting non-Hermitian physics.

The standard solution to break reciprocity is by using a static
magnetic field bias. The magnetic field is odd under the
TR operation and thereby may give rise to a gyrotropic
nonreciprocal permeability in ferrimagnetic materials
\cite{pozar_microwave_2011,adam_ferrite_2002} or a gyroelectric
nonreciprocal permittivity in magnetized plasmas
\cite{krall_principles_1973,pitaevskii_physical_1981}. Such material
platforms are extensively used to realize optical isolators and
circulators \cite{adam_ferrite_2002,dotsch_applications_2005}, and
due to their topological properties \cite{silveirinha_chern_2015}
may enable ``one-way'' propagation free of back-scattering
\cite{haldane_possible_2008,raghu_analogs_2008,yu_one_2008,wang_observation_2009,davoyan_theory_2013,lin_unidirectional_2013,lu_topological_2014,gangaraj_topologically_2017,khanikaev_two-dimensional_2017,fernandes_topological_2019,lannebere_link_2018,lannebere_photonic_2019}.
However, the necessity of an external magnetic field is a major
drawback for the integration of such components on a chip.
Alternative magnetless solutions have been investigated in recent
years. Such solutions can be sub-divided into two classes, depending
if they preserve or not the linearity of Maxwell's equations.

For the first class, the material response is linear under normal
conditions of operation. Such solutions include  time-variant
 systems
\cite{yu_complete_2009,manipatruni_optical_2009,kamal_noiseless_2011,huang_complete_2011,
fang_photonic_2012,sounas_giant_2013,wang_optical_2013,
hua_demonstration_2016,sounas_non-reciprocal_2017,ruesink_nonreciprocity_2016,miri_optical_2017,ruesink_optical_2018,galiffi_broadband_2019,mock_magnet-free_2019,duggan_optically_2019},
 systems with moving
parts \cite{horsley_optical_2013,fleury_sound_2014,lannebere_wave_2016},
systems with drifting electrons
\cite{borgnia_quasi-relativistic_2015,duppen_current-induced_2016,morgado_drift_2018,wenger_current-controlled_2018,bliokh_electric_2018,morgado_nonlocal_2020,morgado_active_2021,zhao_efficient_2021,dong_fizeau_2021},
and non-Hermitian platforms, e.g. PT-symmetric systems
\cite{makris_beam_2008,ruter_observation_2010,lin_experimental_2012},
%
%Professor: Here I deleted some references since they were also nonlinear (PT symmetric and nonlinear). I also put out the article by Fan, since I do not believe it is PT symmetric, am I wrong?
% Sylvain: I edited slightly the citation of Ref. 49, please check!
%Professor:I agree
%
 active electronic systems
\cite{kodera_artificial_2011,wang_gyrotropic_2012,kodera_magnetless_2013}
or optically pumped systems \cite{buddhiraju_nonreciprocal_2020}. It
is relevant to point out that  time-modulations, the drift current
bias or the velocity bias, all imply an explicitly broken
TR symmetry. In contrast, as highlighted recently in Ref.
\cite{buddhiraju_nonreciprocal_2020}, a non-Hermitian response can
be compatible with the TR symmetry, but yet break the
reciprocity. All the solutions in this first class require some
external bias of the system.

The second class is formed by systems that exploit \emph{dynamic}
nonlinear effects and that have been implemented in photonic crystals
\cite{scalora_photonic_1994,tocci_thinfilm_1995,gallo_all-optical_1999,gallo_all-optical_2001,mingaleev_nonlinear_2002,zhou_all-optical_2006,lin_high_2008,miroshnichenko_reversible_2010,shadrivov_electromagnetic_2011,lepri_asymmetric_2011,zhukovsky_all-optical_2011} or using Fano resonances \cite{ding_ultrahigh_2012,xu_reconfigurable_2014,yu_nonreciprocal_2015,sounas_broadband_2018},  PT-symmetry\cite{sukhorukov_nonlinear_2010,ramezani_unidirectional_2010,peng_paritytime_2014} or other mechanisms\cite{fan_all-silicon_2012,fan_silicon_2013,anand_optical_2013,mahmoud_all-passive_2015,fernandes_asymmetric_2018}. These systems are typically self-biased by the incoming wave. Thus, they require input signals with very large power, and generally speaking they cannot provide a robust optical isolation \cite{shi_limitations_2015,khanikaev_nonlinear_2015,fernandes_asymmetric_2018}.

Here, inspired by the physics of transitors, we unveil a different opportunity to generate a strongly nonreciprocal and non-Hermitian \emph{linearized} material response, which relies on the combination of a static electric bias with nonlinearities. 
Crucially, even though the material nonlinearity
is essential to break the reciprocity, in our system the effective
material response to weak dynamical signals can be assumed linear.
This property may be understood with an analogy with semiconductor
transistors, which are nonlinear devices biased by static fields,
and then behave as linear systems (e.g., amplifiers) for relatively
weak dynamical signals.

It is curious to note that the nonreciprocal responses of other
well-known systems with a broken TR symmetry also rely on nonlinear
effects that are linearized around a biasing point for small
amplitude signals. For example the gyrotropic permeability of
ferrimagnetic materials and the gyroelectric permittivity of
magnetized plasmas result from the linearization of the magnetic
torque and of the Lorentz force, respectively
\cite{pozar_microwave_2011}. Similarly, the nonreciprocal response
of materials with a drift current results from the linearization of
the Boltzmann equation
\cite{krall_principles_1973,pitaevskii_physical_1981,sydoruk_terahertz_2010,morgado_negative_2017,morgado_reply_2019}.
Furthermore, the linearization of optomechanical interactions is at
the origin of the nonreciprocity reported in Refs.
\cite{huang_complete_2011,poulton_design_2012,hafezi_optomechanically_2012}.

Here, inspired by this feature and by the operation of standard
transistors, we propose a new paradigm to obtain a nonreciprocal
linearized response in a bulk nonlinear material, which combines the
benefits of the linearized systems cited above with the simplicity
of the electric field bias, providing a unique platform for the
study of non-Hermitian and non-reciprocal wave phenomena.

%remaining TR symmetric, similar to the solution recently reported
%in \cite{buddhiraju_nonreciprocal_2020}.

%Metal Oxide Semiconductors Field Effect Transistors (MOSFETs
%\cite{sze_physics_2021,pozar_microwave_2011})
%\section{Operation principle of the Mosfet metamaterial}

\begin{figure*}[!ht]
\centering
\includegraphics[width=\linewidth]{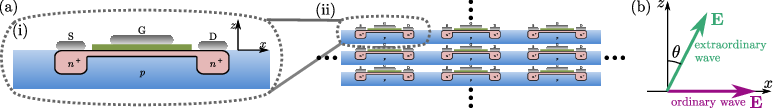}
       \caption{(a)(i) Sketch of a MOSFET transistor. (ii) Illustration of a metamaterial formed by a periodic array of MOSFETs. (b) Geometrical relation between the $E$-fields of the plane wave modes in the MOSFET-metamaterial [Eq. \eqref{E:permittivity_MOSFET_MTM}] for propagation along $y$.}
\label{fig:sketch_mosfet}
\end{figure*}

As a starting point, consider a standard MOSFET transistor, as
depicted in Fig. \ref{fig:sketch_mosfet}(a)-(i). In a MOSFET, the
voltage applied on the gate controls the height of the channel that
connects the drain and the source
\cite{sze_physics_2021,pozar_microwave_2011}. Thus, the electric
field along the $z$-direction controls the ``polarizability'' of the
system along the $x$-direction.
Let us imagine a metamaterial formed by many structural unities
identical to the MOSFET organized in a lattice (Fig.
\ref{fig:sketch_mosfet}(a)-(ii)). Hereafter we refer to this medium as
the MOSFET-metamaterial. We note that metamaterials loaded with
transistors have been discussed in seminal works by Caloz and others
\cite{kodera_artificial_2011,wang_gyrotropic_2012,kodera_magnetless_2013}
in a different context.

The material constitutive relation that mimics the response of a
MOSFET transistor is of the form $ \vec{P} =\eps_0
\db{\chi}(\vec{E}) \cdot \vec{E}$ with
\begin{align}  \label{E:non_linear_susceptibility}
 \db{\chi}(\vec{E})&= \begin{pmatrix} \chi_{xx}(E_z) & 0 & 0 \\ 0 & \chi_{yy} & 0 \\ 0 & 0 & \chi_{zz}\end{pmatrix}.
\end{align}
Here, $\vec{P}$ is the polarization vector in the metamaterial and
$\vec{E}$ is the electric field. The susceptibility $\chi_{xx}(E_z)$
depends on the field strength along $z$, in the same manner as the
MOSFET impedance along the drain-to-source channel ($x$-direction)
depends on the gate voltage ($z$-component of the electric field).
Thus, the metamaterial is nonlinear. The polarizabilities
$\chi_{yy}$ and $\chi_{zz}$ are assumed independent of the field
strength. For simplicity we neglect material dispersion so that the susceptibility is frequency independent and real-valued. The dispersive effects could be modelled by noting that the material response is determined by a set of nonlinear differential equations, which could be linearized using a procedure analogous to what is described below. The main impact of material dispersion is that it usually determines a frequency cutoff beyond which the nonlinear response becomes too weak, and, in addition, the permittivity components may become complex-valued.

Let us suppose that such hypothetical metamaterial is biased with
some static electric field in the $xoz$ plane $\vec{E}_0=E_{0x}
\hat{\vec{x}} + E_{0z} \hat{\vec{z}}$. For small field variations
($\delta \vec{E}$) around the biasing point ($\vec{E}=\vec{E}_0 +
\delta \vec{E}$) the electric response can be linearized as,
\begin{align}
\vec{P} &\approx \eps_0 \db{\chi}(\vec{E}_0) \cdot \vec{E}_0 +
\eps_0 \sum_{i=x,y,z} \partial_{E_i} \left.\left[ \db{\chi}(\vec{E})
\cdot \vec{E}\right]\right|_{\vec{E}=\vec{E}_0} \delta E_i
\end{align}
In the above, $\partial_{E_i}\equiv \partial/\partial E_i$ with
$i=x,y,z$, represents a derivative with respect to the electric
field. The induced polarization is $\vec{P}=\vec{P}_0 + \delta
\vec{P}$, with $\delta \vec{P}=\eps_0 \sum_{i=x,y,z}
\partial_{E_i} \left.\left[ \db{\chi}(\vec{E}) \cdot
\vec{E}\right]\right|_{\vec{E}=\vec{E}_0} \delta E_i $ the
linearized signal response. Thus, for sufficiently weak signals, the
dynamical parts of the polarization and electric field vectors are
related through a linear relation of the form $\delta \vec{P}=\eps_0
\db{\chi}_\text{lin} \cdot \delta \vec{E}$ where
$\db{\chi}_\text{lin}$ is the effective (linearized) material
susceptibility. It can be written explicitly as:
\begin{align}   \label{E:susceptibilityMOSFET}
\db{\chi}_\text{lin} &=\sum_{i=x,y,z}
\partial_{E_i} \left.\left[ \db{\chi}(\vec{E}) \cdot
\vec{E}\right]\right|_{\vec{E}=\vec{E}_0} \otimes {{\bf{\hat u}}_i},
\end{align}
with ${{\bf{\hat u}}_i}$ a unit vector directed along the
\emph{i}-th direction, and $\otimes$ represents the tensor product
of two vectors. For the particular
model in Eq. \eqref{E:non_linear_susceptibility}, one finds that
\begin{align}\label{E:chi_linear}
\db{\chi}_\text{lin}=\begin{pmatrix} \chi_{xx}(E_{0z}) & 0 & \left.
\partial_{E_z} \chi_{xx}\right|_{\vec{E}=\vec{E}_0}
E_{0x} \\ 0 & \chi_{yy} & 0 \\ 0 & 0 & \chi_{zz}\end{pmatrix}.
\end{align}
In general, the linearized response is nonreciprocal because
$\db{\chi}_\text{lin}\neq \db{\chi}_\text{lin}^T$ (the superscript
$T$ represents the matrix transpose). Thus, our proposal establishes a novel paradigm to
have nonreciprocity with an
electric bias. A nonreciprocal response
requires that the bias static field has an $x$-component
($E_{0x}\neq0$), i.e. a component along the drain-to-source
direction in the MOSFET transistor analogue. The field component
$E_{0z}$ may be zero when $\left. \partial_{E_z}
\chi_{xx}\right|_{E_z=0} \neq 0$.

It is rather curious that the microscopic susceptibility is a
symmetric tensor (Eq.\eqref{E:non_linear_susceptibility}), but the linearized susceptibility (Eq.\eqref{E:chi_linear}) is not. Furthermore, for the considered model, the linearization preserves the TR symmetry of the system.  We note in passing that Ref.
\cite{buddhiraju_nonreciprocal_2020} studied related (but not
equivalent) non-Hermitian material responses obtained using optical
pumping. In general, a non-Hermitian material response must be rooted in some nonequilibrium process
that can extract energy from either the stored electrostatic field or from the DC generator. The latter case is feasible with a drift current, analogous to a MOSFET transistor. Note that in our scheme the drift velocity can be a tiny fraction of the wave velocity, different from other solutions studied previously that rely on materials with large mobility \cite{borgnia_quasi-relativistic_2015,duppen_current-induced_2016,morgado_drift_2018,wenger_current-controlled_2018,bliokh_electric_2018,morgado_nonlocal_2020,morgado_active_2021,zhao_efficient_2021,dong_fizeau_2021}.

Even though our starting point was an analogy with transistors,
related nonreciprocal responses can in principle be engineered using
naturally available nonlinear materials. For example, related
nonlinearities may naturally occur in crystals without inversion
symmetry \cite{yariv_optical_1984} and may also be
engineered in semiconductor superlattices
\cite{rosencher_quantum_1996}. It is however essential that the material is operated in  nonequilibrium (e.g., with carrier injection), as in equilibrium the Kleinman symmetry forces the linearized response to be reciprocal \cite{kleinman_nonlinear_1962,boyd_nonlinear_2008}.

Since the electric displacement vector is $\vec{D}=
\eps_0\vec{E}+\vec{P}$, the linearized relative permittivity of the
metamaterial is $\db{\eps}=\vec{1}_{3\times3}+\db{\chi}_\text{lin}$
. Thus, for a MOSFET-metamaterial the permittivity tensor is of the
form:
\begin{align}  \label{E:permittivity_MOSFET_MTM}
\db{\eps} =\begin{pmatrix} \eps_{xx}  & 0 & \eps_{xz}  \\ 0 & \eps_{yy}  & 0 \\ 0 & 0 & \eps_{zz}\end{pmatrix}.
\end{align}
This formula confirms that when the permittivity tensor is
real-valued and frequency independent, the linearized response
remains TR invariant ($\db{\eps}^\ast=\db{\eps}$), even
though it is nonreciprocal ($\db{\eps}^T\neq\db{\eps}$)
\cite{altman_reciprocity_2011}. Furthermore, the metamaterial is
non-Hermitian ($\db{\eps}^\dagger\neq\db{\eps}$), which indicates
that it can either absorb or generate energy.

%%%%%%%%%%%%%%%%%%%%%%%%%%%%%%%%%%%%%%%%%%%%%%%%%%%%%%%%%%%%%%%%%%%%%%%%%%%%%%%%%%%%%%%%%%
%%%%%%%%%%%%%%%%%%%%%%%%%%%%%%%%%%%%%%%%%%%%%%%%%%%%%%%%%%%%%%%%%%%%%%%%%%%%%%%%%%%%%%%%%%

%\section{Plane wave propagation and power beating}

For simplicity, in the remainder of this Letter, we focus on the
material response of the idealized MOSFET-metamaterial
\eqref{E:permittivity_MOSFET_MTM}. We are interested only in weak
signals described by the linearized response, and thus hereafter, to
keep the notations short, we drop the $\delta$ symbol ($\delta
\vec{E} \to \vec{E}$), and denote the dynamic electric field simply
by $\vec{E}$.

Next, we characterize the plane wave modes in the metamaterial. The
wave equation is of the form  $\nabla \times \nabla \times  \vec{E}
=  \frac{\omega^2}{c^2} \db{\eps}   \cdot  \vec{E}$, where $\omega$ is the angular frequency and $c$ the speed of light in vacuum. Plane wave solutions are of
the form $\vec{E}=\vec{A}_0\e{i\vec{k}\cdot \vec{r}}$ with
$\vec{A}_0$ a constant complex vector, $\vec{k}$ the wave vector and $\vec{r}=x\hat{\vec{x}}+y\hat{\vec{y}}+z\hat{\vec{z}}$ the position vector.
We restrict our attention to propagation along the $y$- direction,
i.e., the direction perpendicular to the Gate-Drain-Source plane
(see Fig. \ref{fig:sketch_mosfet}(a)), so that $\vec{k}=k
\hat{\vec{y}}$.
%For now, we also assume that the permittivity tensor is real-valued and frequency
%independent.
In this case, the solutions of the homogeneous wave equation are
such that:
 \begin{subequations}\label{E:modes}
\begin{align}
  &\vec{k}= \frac{\omega}{c}\sqrt{\eps_{xx}} \hat{\vec{y}}\equiv \vec{k}_o, \qquad   \vec{E} \sim \hat{\vec{x}} & \text{(ordinary wave)} \\
 &\vec{k}= \frac{\omega}{c}\sqrt{\eps_{zz}} \hat{\vec{y}}\equiv \vec{k}_e, \qquad   \vec{E} \sim \frac{\eps_{xz}}{\eps_{zz}-\eps_{xx}} \hat{\vec{x}} + \hat{\vec{z}} & \text{(extraordinary wave)}
 \end{align}
\end{subequations}
Remarkably, due to the non-Hermitian response for a non-zero
cross-coupling coefficient $\eps_{xz}$, the ordinary and
extraordinary waves are not orthogonal
\cite{cerjan_achieving_2017,buddhiraju_nonreciprocal_2020}. The tilt
of the extraordinary wave electric field with respect to the
direction perpendicular to the ordinary wave is determined by an
angle $\theta$ such that
$\tan(\theta)=\frac{\eps_{xz}}{\eps_{zz}-\eps_{xx}}$ (see Fig.
\ref{fig:sketch_mosfet}(b)).
% %
% \begin{figure}[!ht]
% \centering
% \includegraphics[width=.35\linewidth]{eigenvectors_Mosfet}
%        \caption{Geometrical relation between the electric fields associated with the plane waves propagating
% along the $y$-direction \eqref{E:modes} in a MOSFET-metamaterial described by a frequency independent permittivity tensor given by Eq. \eqref{E:permittivity_MOSFET_MTM}.}
% \label{fig:eigenvectors_Mosfet}
% \end{figure}
% %
Interestingly, $\theta$ determines the strength of the nonreciprocal
effects. It may differ significantly from zero even for a small
cross-coupling coefficient $\eps_{xz}$ provided the values of
$\eps_{xx}$ and $\eps_{zz}$ are sufficiently close. The limiting
case $\eps_{xx}=\eps_{zz}$ corresponds to an exceptional point where
the two waves
 \eqref{E:modes} coalesce into a single linear polarization \cite{buddhiraju_nonreciprocal_2020} and will not be further considered here.

\begin{figure}[!ht]
\centering
\includegraphics[width=0.65\linewidth]{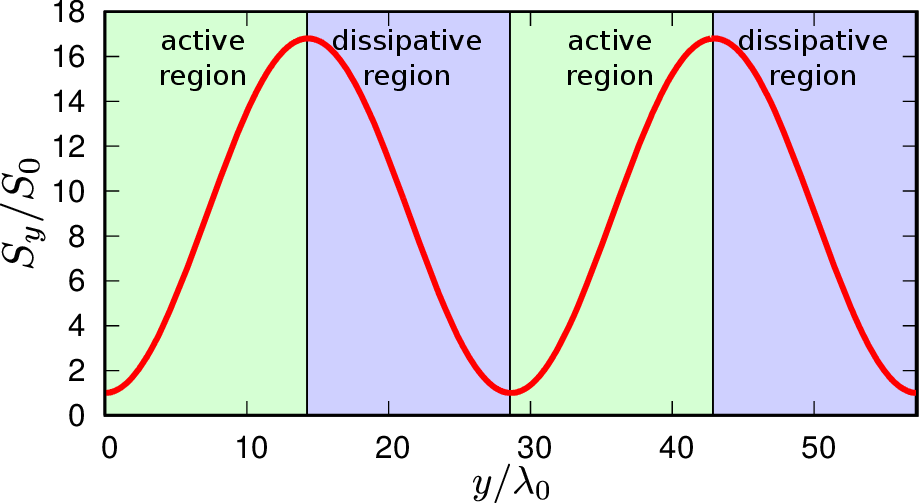}
       \caption{Normalized Poynting vector  as a function of the propagation distance $y$ normalized to the vacuum wavelength $\lambda_0=2\pi c/\omega$.
The parameters are $\eps_{xx}=2$, $\eps_{zz}=2.1$, $\eps_{xz}=0.2$, $A_e=1$ and $A_o=- A_e \frac{\eps_{xz}}{\eps_{zz}-\eps_{xx}}$.
The regions shaded in green correspond to active regions, whereas the regions shaded in purple correspond to dissipative regions.}
\label{fig:power_beating}
\end{figure}

Consider a superposition of ordinary and extraordinary waves
propagating along $+y$:
\begin{align} \label{E:superposition_E}
\vec{E}&=  A_o  \e{i k_o y} \hat{\vec{x}} + A_e \left(   \frac{\eps_{xz}}{\eps_{zz}-\eps_{xx}} \hat{\vec{x}} +\hat{\vec{z}} \right) \e{i k_e y}
\end{align}
The corresponding magnetic field is given by $\vec{H}=
\frac{-i}{\omega \mu_0}\hat{\vec{y}}  \times \partial_y \vec{E}$
(with $\partial_y \equiv \partial/\partial y$).
% is
 %
%\begin{align} \label{E:superposition_H}
%\vec{H}  &= -    A_o \frac{k_o}{\omega \mu_0}  \hat{\vec{z}} \e{i k_o y} + \frac{k_e}{\omega \mu_0}A_e        \left( \hat{\vec{x}}    -\frac{\eps_{xz}}{\eps_{zz}-\eps_{xx}} \hat{\vec{z}} \right) \e{i k_e y}.
%\end{align}
%
After some algebra, one may show that the time-averaged Poynting vector $\vec{S}=\frac{1}{2} \mathrm{Re}\left\{\vec{E} \times \vec{H}^\ast \right\}$ in the material is:
\begin{align} \label{E:Poynting_analytic}
\vec{S} &= \frac{1}{2}\frac{\sqrt{\eps_0}}{\sqrt{\mu_0}} \left( |A_o|^2    \sqrt{\eps_{xx}}  + |A_e|^2 \sqrt{\eps_{zz}}  \left(  1 + \left| \frac{\eps_{xz}}{\eps_{zz}-\eps_{xx}} \right|^2 \right)  +\frac{1}{\sqrt{\eps_{zz}} - \sqrt{\eps_{xx}}}   \mathrm{Re}\left\{ \eps_{xz}  A_e A_o^\ast  \e{i (k_e - k_o) y}\right\}  \right)\hat{\vec{y}}
\end{align}
The first two terms of this expression represent the power transported by the ordinary and the
extraordinary waves alone, respectively. Remarkably, provided the cross-coupling coefficient $\eps_{xz}$ does not vanish, there is a third term that is responsible
 for a periodic spatial variation of the power flux. This third term describes a \textit{power beating} with spatial frequency $k_e-k_o$ as illustrated with a numerical example
 in Fig. \ref{fig:power_beating}.
Thus, when $\eps_{xz}\neq 0$, the two modes cease to transport the
power independently in the material. This is a rather unique result,
as for any Hermitian system the electromagnetic modes always
transport power independently. In fact, if the crossed term of the
Poynting vector does not vanish, the power flux forcibly depends on
the propagation distance (as in Fig. \ref{fig:power_beating}). This is only possible in
non-energy conserving (non-Hermitian) systems, e.g., systems with
loss or gain ($\db{\eps}\neq\db{\eps}^\dagger$).

Curiously, the interplay of the static bias field with the material
nonlinearity may lead to either dissipation or gain. Thus, our
metamaterial may behave as either ``lossy'' or ``gainy'', depending
on the relative phase between the two propagating modes. In
different words, the material is neither a standard lossy material
nor a standard gainy material, but rather exhibits a dual-type
response.
 In particular, depending on the relative phase between the
ordinary and extraordinary waves, the MOSFET-metamaterial alternates
between active and dissipative regions (see the supplementary
information \cite{supplemental_material}). For the particular model considered here, the energy exchange
between the wave and the medium is controlled by a polarization
current induced by the static electric field through the
nonlinearity \cite{supplemental_material}. 
 The power amplification is
proportional to the magnitude of the cross-coupling coefficient and
can be made rather large if the values of $\eps_{xx}$ and
$\eps_{zz}$ are sufficiently close. Similar power beatings may occur in
other non-Hermitian platforms (e.g.,
\cite{buddhiraju_nonreciprocal_2020}).
%Sylvain: can we find other examples in PT symmetric systems?
%Professor: from I have seen, no.

%\section{MOSFET-metamaterial isolator}

\begin{figure*}[!ht]
\centering
\includegraphics[width=\linewidth]{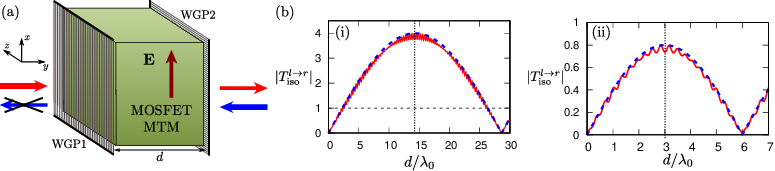}
       \caption{(a) Electromagnetic isolator based on an electrically biased MOSFET-metamaterial placed in between two orthogonal linear polarizers.
To ease the vizualization the polarizers are represented as wire grids, WGP1 and WGP2, that absorb electric fields parallel to the directions of the wires, i.e.,
the $x$ and $z$ directions, respectively. The propagation from right-to-left is forbidden.
The direction of the electric bias $\vec{E}_0$ is also shown. (b) Transmission from left to right $\left|T_\text{iso}^{l\to r}\right|$ as a function of the isolator thickness $d$ normalized to the free-space wavelength. The parameters are  $\eps_{xx}=2$, $\eps_{xz}=0.2$ and (i) $\eps_{zz}=2.1$ and (ii) $\eps_{zz}=2.5$. The red line is the exact solution whereas the blue dashed line is the approximate solution given by Eq.\eqref{E:transmission_approched_isolator}. The vertical dotted lines mark the positions of $d_\text{max}$.}
\label{fig:isolator}
\end{figure*}

Next, we  present a design of an electromagnetic isolator. The proposed system is depicted in Fig.
\ref{fig:isolator}(a) and is based on the non-orthogonality of the two
eigenvectors of the bulk material.
It consists of a MOSFET-metamaterial slab of thickness $d$ placed in
between two orthogonal linear polarizers. The linear polarizers are
supposed to fully absorb the electric field component parallel to
some axis and let the orthogonal component pass through them
unchanged.

We introduce the
transmission matrix $\db{T}$
%$\db{T}\equiv \begin{pmatrix}
%     T_{11}   & T_{12}\\ T_{21} &T_{22}\end{pmatrix}$
that relates (in the absence of the polarizing grids) the transverse components
of the incident $\vec{E}_\text{t}^\text{inc}=\left(E_{x}^\text{inc}
\; E_{z}^\text{inc} \right)^T$ and transmitted
$\vec{E}_\text{t}^\text{tr}=\left(E_{x}^\text{tr} \; E_{z}^\text{tr}
\right)^T$ electric fields as $
 \vec{E}_\text{t}^\text{tr}
=\db{T} \cdot \vec{E}_\text{t}^\text{inc} $.
%
%\begin{align} \label{E:def_matrix_T}
% \vec{E}_\text{t}^\text{tr}
%=\db{T} \cdot \vec{E}_\text{t}^\text{inc}
%\end{align}
%
In the supplementary information \cite{supplemental_material}, it is
shown that for normal incidence ($\vec{k}=k\hat{\vec{y}}$ with $k=\omega/c$) the
transmission matrix for a material slab with thickness $d$ is
\begin{align}  \label{E:T_matrix_MOSFET_MTM}
\db{T}      =\begin{pmatrix}
     \gamma_o   & \frac{\eps_{xz}}{ \eps_{xx}-\eps_{zz}} \left( \gamma_o-\gamma_e\right)\\
 0 &   \gamma_e
\end{pmatrix},
\end{align}
where $\gamma_j=\left( \cos (k_j d)-\frac{i (k_j^2+k^2) }{2
k_jk}\sin (k_j d)\right)^{-1}$ with $j=o,\,e$. The transmission
matrix lacks transpose symmetry and is independent of the direction
of propagation of the incoming wave. The matrix has a single nonzero anti-diagonal element, and hence the slab can generate an outgoing wave with $E_x$ component from an incoming wave polarized along $z$, but not the converse.

Let us now analyze the response of the proposed isolator. 
For an incident wave propagating from left-to-right the polarizing grid
WPG1 ensures that the field that reaches the metamaterial is
oriented along the $z$ direction with an amplitude
$E_{z}^\text{inc}$. Due to the cross-coupling term $\eps_{xz}$, the
propagation in the metamaterial may induce an electric field
component along the $x$-direction. This $x$-component of the field
can go through the grid WPG2, generating the left to right output
signal. 
% From Eq.\eqref{E:T_matrix_MOSFET_MTM} the left to right
% transmission coefficient of the isolator $T_\text{iso}^{l\to
% r}=T_{12}$  is precisely (assuming ideal polarizing grids):
% %
% \begin{align} \label{E:transmission_isolator}
% \left|T_\text{iso}^{l\to r}\right|= \left| \frac{\eps_{xz}}{ \eps_{xx}-\eps_{zz}} \left( \gamma_o-\gamma_e\right) \right|
% \end{align}
% %
% Note that the transmission coefficient is non-zero only when
% $\eps_{xz}\neq0$. 
In contrast, an incident wave propagating from
right-to-left can enter into the MOSFET-metamaterial only if the incident
electric field has a component along $x$. In this case, the metamaterial
does not generate any cross-polarization component, and thereby the
incoming wave is fully absorbed at the output polarizing grid. 
Therefore, the wave propagation is allowed only in a specific
(left-to-right) direction.\\
This is confirmed by Eq.\eqref{E:T_matrix_MOSFET_MTM} that reveals that
the right to left transmission coefficient $T_\text{iso}^{r\to
l}=T_{21}$ of the isolator is exactly
zero. 
% %
% \begin{figure}[!ht]
% \centering
% \includegraphics[width=0.8\linewidth]{transmission_isolator}
%        \caption{Transmission from left to right $\left|T_\text{iso}^{l\to r}\right|$ for the isolator of Fig. \ref{fig:isolator} as a function of the thickness $d$
%         normalized to the vacuum wavelength. The parameters are $k=k_0$, $\eps_{xx}=2$, $\eps_{xz}=0.2$ and (a)
%         $\eps_{zz}=2.1$ and (b)
%         $\eps_{zz}=2.5$. The red line is the exact solution given by $T_{12}$ whereas the blue dashed line is
%         the approximate solution given by
%         Eq.\eqref{E:transmission_approched_isolator}. The vertical dotted
%         lines mark the positions of $d_\text{max}$.  }
% \label{fig:transmission_isolator}
% \end{figure}
% %
The left-to-right transmission $\left|T_\text{iso}^{l\to r}\right|=T_{12}$  is depicted in Fig. \ref{fig:isolator}(b) as a function of the isolator
thickness $d$ for different material parameters. As seen, a longer
propagation distance typically yields a stronger output signal, up  to some threshold value beyond which an oscillatory behavior is
observed.
By neglecting the reflections at the air-metamaterial interfaces, one finds that \cite{supplemental_material}:
\begin{align} \label{E:transmission_approched_isolator}
\left|T_\text{iso}^{l\to r}\right| \approx \left|  2  \frac{\eps_{xz}}{\eps_{zz}-\eps_{xx}}   \sin\left(\frac{k_e -k_o}{2}d\right)      \right|.
\end{align}
Hence, the transmission is maximized when $\frac{k_e
-k_o}{2}d=\frac{\pi}{2}$, i.e. for a thickness $d =
d_\text{max}\equiv\frac{\lambda_0}{2(\sqrt{\eps_{zz}}-\sqrt{\eps_{xx}})}$.
The transmission maximum is $\left|  2  \tan(\theta)\right| $ and is
controlled by the angle $\theta$ and by the non-orthogonality of the
eigenvectors (see Fig. \ref{fig:sketch_mosfet}(b)). The
transmission level can in principle be much greater than unity due
to the active response of the metamaterial (see
Fig. \ref{fig:isolator}(b)-(i)).

For a fixed value of $\eps_{xz}$, a larger detuning of the diagonal
elements $|\eps_{xx}-\eps_{zz}|$ leads to a decrease in
$d_\text{max}$, but at the same time the transmission level drops
down. This is illustrated in Fig. \ref{fig:isolator}(b)
where the strong amplification obtained in (i) after a propagation
of around $15\lambda_0$ becomes smaller than unity but the maximum
is reached only after $3\lambda_0$.

% The behavior of $\left|T_\text{iso}^{l\to r}\right|$ is
% closely linked to the behavior of the Poynting vector in the
% metamaterial. Under the same approximations (no reflections), the Poynting vector
% (Eq.\eqref{E:Poynting_analytic}) in the isolator for left-to-right
% propagation is
% %
% \begin{align}
% \vec{S} &= \frac{|A_e|^2}{2}\frac{\sqrt{\eps_0}}{\sqrt{\mu_0}} \left( \sqrt{\eps_{zz}} + \left| \frac{\eps_{xz}}{\eps_{zz}-\eps_{xx}} \right|^2( \sqrt{\eps_{xx}} +\sqrt{\eps_{zz}})\left(1- \cos\left[ (k_e - k_o) y\right] \right)  \right)\hat{\vec{y}}
% \end{align}
% %
% The minimum value occurs at $y=0$ and, as $y$ increases, the
% Poynting vector increases monotonically leading to an enhancement of
% the power transported in the metamaterial (see Fig.
% \ref{fig:power_beating} (a)). Interestingly the first maximum occurs
% at the same distance $d=d_\text{max}$ where the transmission is
% maximized (compare Figs. \ref{fig:power_beating} (a) and
% \ref{fig:transmission_isolator} (a)).
% % Thus, the cross-polarization conversion is maximized when the Poynting vector amplitude also is.
For a device with small $d$ Eq. \eqref{E:transmission_approched_isolator} becomes $\left|T_\text{iso}^{l\to r}\right| \approx   \frac{2\pi
d}{\lambda_0} \left|  \frac{\eps_{xz}}{\sqrt{\eps_{zz}}
+\sqrt{\eps_{xx}}}         \right|$.
Hence, for a small thickness the amplification factor is bounded by
the amplitude of $\eps_{xz}$. In contrast, for a large thickness
this limitation disappears, because the distributed gain is controlled by
$\frac{\eps_{xz}}{\eps_{zz}-\eps_{xx}}$, and thus can be arbitrarily
large.

It is relevant to point out that even though the idealized metamaterial is TR invariant, the presence of the dissipative polarizing
grids breaks the TR symmetry of the system (Fig. \ref{fig:isolator}(a)).
In fact, it is well known that dissipation is an absolutely essential ingredient 
to realize an electromagnetic isolator. Moreover, our system is not affected
by any of the problems that limit the performance of systems that
exploit dynamic nonlinearities
\cite{shi_limitations_2015,khanikaev_nonlinear_2015,fernandes_asymmetric_2018}.
In principle our isolator can be implemented using any linearized
material response with non-orthogonal eigenvectors.

In summary, we introduced a new and robust mechanism to break the
Lorentz reciprocity using the combination of a static electric bias
and material nonlinearities. Starting from an analogy with the
operation of a semiconductor transistor, we showed that the
linearization of a nonlinear material response, can lead to a
non-symmetric and non-Hermitian permittivity response. We studied in
detail the wave propagation in an idealized MOSFET-metamaterial,
showing that the non-Hermitian nature of the wave-matter
interactions opens many exciting opportunities and can lead to a
dual material behavior, where regimes of absorption alternate with
regimes of gain in the same physical system. Furthermore, we
suggested  a design of an electromagnetic isolator using the proposed
material, whose performance may in principle  surpass that of
standard Faraday isolators. We envision
that related platforms can be engineered as metamaterials or,
alternatively, can be implemented using naturally available materials in nonequilibrium situations. The prospects are especially promising up to THz frequencies, both in metamaterial \cite{mei_first_2015} and in natural material realizations. Note that the nonlinear response of naturally available materials remains strong and fast enough up to the longitudinal optical phonon resonance \cite{wemple_electrooptical_1972}.

\section*{Acknowledgments}
This work was partially funded by Huawei Technologies, by the
Institution of Engineering and Technology (IET) under the A F Harvey
Research Prize 2018, by the Simons Foundation under the award 733700
(Simons Collaboration in Mathematics and Physics, "Harnessing
Universal Symmetry Concepts for Extreme Wave Phenomena") and by
Instituto de Telecomunica\c{c}\~{o}es under Project N$^\circ$.
UID/EEA/50008/2020. S.L. acknowledges FCT and IT-Coimbra for
the research financial support with reference DL
57/2016/CP1353/CT000. D.E.F. acknowledges support by FCT,
POCH, and the cofinancing of Fundo Social Europeu under the
fellowship SFRH/BPD/116525/2016. T. A. M. acknowledges FCT for research financial support with reference CEECIND/04530/2017 under the CEEC Individual 2017, and IT-Coimbra for the contract as an assistant researcher with reference CT/N$^\circ$. 004/2019-F00069.

\section*{References}
% \bibliographystyle{apsrev4-1}
% \bibliography{Biblio_CLEAN}


\end{document}

\title{Supplemental Material for the Manuscript: \\``Nonreciprocal and non-Hermitian material response inspired by semiconductor transistors''}
\author{Sylvain Lanneb\`{e}re\textsuperscript{1}}
\author{David E. Fernandes\textsuperscript{1}}
\author{Tiago A. Morgado\textsuperscript{1}}
\author{M\'{a}rio G. Silveirinha\textsuperscript{2}}
\email{To whom correspondence should be addressed:
mario.silveirinha@co.it.pt}
 \affiliation{\textsuperscript{1}
Instituto de Telecomunica\c{c}\~{o}es and Department of Electrical Engineering, University of Coimbra, 3030-290 Coimbra, Portugal}
\affiliation{\textsuperscript{2}University of Lisbon -- Instituto
Superior T\'ecnico and Instituto de Telecomunica\c{c}\~{o}es, Department of Electrical Engineering, 1049-001
Lisboa, Portugal}

\begin{abstract}\normalsize \it \vspace{0.2cm}
The supplemental material reports (I) the transmission matrix of a MOSFET-metamaterial for normal incidence, (II) the transmission coefficients of the isolator, (III) and (IV) the study of the energy exchanged between the material and the wave, based on macroscopic and microscopic considerations, respectively. 
\end{abstract}

\maketitle

\section{Transmission matrix for a MOSFET-metamaterial slab}
The transmission matrix for a metamaterial slab of thickness $d$
standing in air can be obtained in a standard way using mode
matching. For the case of normal incidence ($\vec{k}=k\hat{\vec{y}}$), using Eq.
(7) of the main text, one can write the
electric fields in each region as
\begin{align} \label{E:E_each_region}
\vec{E}=
   \begin{cases}
\left( \vec{1}_{2\times2}\e{iky}+ \db{R}  \e{-iky} \right) \cdot \vec{E}_\text{t}^\text{inc}
   & y<0 \\
 \vec{E}_{o,0} \left[A_{o,+}  \e{i k_o y} + A_{o,-}  \e{-i k_o y} \right] +  \vec{E}_{e,0} \left[ A_{e,+}\e{i k_e y}  +  A_{e,-} \e{-i k_e y} \right]   & 0 \leq y\leq d \\
 \db{T} \cdot \vec{E}_\text{t}^\text{inc} \e{ik(y-d)}  &  y > d \\
   \end{cases}
\end{align}
where $A_{o,\pm}$, $A_{e,\pm}$ are the amplitude of the left/right
propagating ordinary and extraordinary waves respectively, $\db{R}$
and $\db{T}$ relate the transverse amplitudes of the reflected and
transmitted fields, and $\vec{E}_{o,0}=   \hat{\vec{x}}$,
$\vec{E}_{e,0}= \frac{\eps_{xz}}{\eps_{zz}-\eps_{xx}} \hat{\vec{x}}
+  \hat{\vec{z}} $ are two vectors proportional to the system
eigenvectors defined in Eq. (6) of the main text.
Similarly, the magnetic field can be written as
%using Eq.\eqref{E:superposition_H} as
%
\begin{align}  \label{E:H_each_region}
\vec{H}=
   \frac{1}{\omega\mu_0} \vec{J} \cdot \begin{cases}
k   \left( \vec{1}_{2\times2}\e{iky}-   \db{R} \e{-iky} \right)\cdot \vec{E}_\text{t}^\text{inc}
   & y<0 \\
  k_o    \vec{E}_{o,0} \left( A_{o,+}   \e{i k_o y} -  A_{o,-}    \e{-i k_o y}\right) +   k_e   \vec{E}_{e,0}\left(  A_{e,+} \e{i k_e y}  -    A_{e,-}   \e{-i k_e y} \right)   & 0 \leq y\leq d \\
 k  \db{T} \cdot \vec{E}_\text{t}^\text{inc} \e{ik(y-d)}  &  y > d \\
   \end{cases}
\end{align}
where $\vec{J}=\begin{pmatrix}0 & -1\\ 1&0\end{pmatrix}$. By
matching the fields in \eqref{E:E_each_region} and
\eqref{E:H_each_region} at $y=0$ and $y=d$ we obtain a system of
linear equations whose solutions give after simplification 
% it is obtained that the
% transmission matrix of a MOSFET-metamaterial slab of thickness $d$ under normal incidence is
% %
% \begin{align}  \label{E:T_matrix_MOSFET_MTM}
% \db{T}      =\begin{pmatrix}
%      \gamma_o   & \frac{\eps_{xz}}{ \eps_{xx}-\eps_{zz}} \left( \gamma_o-\gamma_e\right)\\
%  0 &   \gamma_e
% \end{pmatrix},
% \end{align}
% %
% where $\gamma_j=\left( \cos (k_j d)-\frac{i (k_j^2+k^2) }{2
% k_jk}\sin (k_j d)\right)^{-1}$ with $j=o,\,e$. The transmission
% matrix lacks transpose symmetry and is independent of the direction
% of propagation of the incoming wave. A single nonzero anti-diagonal
% element implies that the medium can generate a $x$-component of the
% field from a $z$-oriented incoming electric field, but not the other
% way around.
equation
(9) of the main text.

\section{Transmission coefficients of the isolator}
Consider the isolator depicted in Fig.3 (a) of the main text. Without loss of generality, the incident field may be assumed polarized along the direction perpendicular to the input polarizing grid (which depends on the direction of arrival of the incoming wave). Assuming ideal polarizing grids, the right to left transmission coefficient of the  isolator is $T_\text{iso}^{r\to l}=T_{21}$, which according to Eq.(9) of the main text is exactly zero. On the other hand, the left to right
transmission coefficient of the isolator $T_\text{iso}^{l\to
r}=T_{12}$  is precisely:
\begin{align} \label{E:transmission_isolator}
\left|T_\text{iso}^{l\to r}\right|= \left| \frac{\eps_{xz}}{ \eps_{xx}-\eps_{zz}} \left( \gamma_o-\gamma_e\right) \right|
\end{align}
Note that the transmission coefficient is non-zero only when
$\eps_{xz}\neq0$.

For propagation along the +$y$ direction (left-to-right), it is
possible to get an analytic estimate of $\left|T_\text{iso}^{l\to
r}\right|$ by neglecting the reflections at the air-metamaterial
interfaces. The incident field polarized along $z$ excites a
superposition of the ordinary and extraordinary waves, that
according to Eq.(7) of the main text %\eqref{E:superposition_E}
satisfy $A_o+A_e\left(
\frac{\eps_{xz}}{\eps_{zz}-\eps_{xx}}\right)=0$. Thereby, the
ordinary and extraordinary wave amplitudes are in phase and the
electric field in the metamaterial is:
\begin{align} \label{E:E_field_isolator}
\vec{E}(y)&= A_e\left( 2i \frac{\eps_{xz}}{\eps_{zz}-\eps_{xx}} \e{i \frac{k_e +k_o}{2}y}   \sin\left(\frac{k_e -k_o}{2}y   \right)  \hat{\vec{x}} +    \hat{\vec{z}} \e{i k_e y}    \right)
\end{align}
We note in passing that typically the electric field in the
metamaterial is elliptically polarized. The transmission coefficient
of the isolator may be approximated by $\left|T_\text{iso}^{l\to
r}\right|\approx |\vec{E}(d) \cdot\hat{\vec{x}}| /|\vec{E}(0)\cdot
\hat{\vec{z}}|$ with $\vec{E}$ given by \eqref{E:E_field_isolator}.
After some simplifications, one obtains equation (10) of the main text.\\
% %
% \begin{align} \label{E:transmission_approched_isolator}
% \left|T_\text{iso}^{l\to r}\right| \approx \left|  2  \frac{\eps_{xz}}{\eps_{zz}-\eps_{xx}}   \sin\left(\frac{k_e -k_o}{2}d\right)      \right|.
% \end{align}
% %
As seen in Fig.3 (b) of the main text, %\ref{fig:transmission_isolator}
this approximation
is quite good and explains the origin of the small amplitude
oscillations observed in the exact transmission curve: they are due
to the standing wave in the metamaterial stemming from the
reflections at the second interface.\\

The behavior of $\left|T_\text{iso}^{l\to r}\right|$ is
closely linked to the behavior of the Poynting vector in the
metamaterial. Under the same approximations (no reflections), the Poynting vector
(Eq.(8) of the main text) in the isolator for left-to-right
propagation is
\begin{align}
\vec{S} &= \frac{|A_e|^2}{2}\frac{\sqrt{\eps_0}}{\sqrt{\mu_0}} \left( \sqrt{\eps_{zz}} + \left| \frac{\eps_{xz}}{\eps_{zz}-\eps_{xx}} \right|^2( \sqrt{\eps_{xx}} +\sqrt{\eps_{zz}})\left(1- \cos\left[ (k_e - k_o) y\right] \right)  \right)\hat{\vec{y}}
\end{align}
The first minimum value occurs at $y=0$ and, as $y$ increases, the
Poynting vector increases monotonically leading to an enhancement of
the power transported in the metamaterial (see Fig.2 of the main text). Interestingly the first maximum occurs
at the same distance $d=d_\text{max}$ where the transmission is
maximized (compare Figs.2 and
3-(b)(i) of the main text).
% Thus, the cross-polarization conversion is maximized when the Poynting vector amplitude also is.

\section{Exchange of energy between the wave and the material: macroscopic approach} \label{sec:power_transferred}

Let us consider a nondispersive lossless material described by the
linear permittivity tensor
\begin{align}  \label{E:asymmetric_permittivity}
\db{\eps} =\begin{pmatrix} \eps_{xx}  & 0 & \eps_{xz}  \\ 0 & \eps_{yy}  & 0 \\ 0 & 0 & \eps_{zz}\end{pmatrix}.
\end{align}
The instantaneous power transferred from the wave to the
metamaterial is \cite{kong_electromagnetic_1986}
\begin{align}  \label{E:power_transferred}
Q= \int \vec{j}(t)\cdot \vec{E}(t)~dV
\end{align}
with $\vec{j}(t) =
\partial_t \vec{P}(t)$ the polarization current.
 In time-harmonic regime the electric field is given by
$\vec{E}(t) =\frac{1}{2}\left(\vec{E}_\omega \e{-i\omega t} +
\vec{E}_\omega^\ast \e{i\omega t}\right)$ and the polarization
vector is $\vec{P}(t) =\eps_0\left( \db{\eps} - \vec{1}_{3\times
3}\right) \cdot \vec{E}(t)$. Then, the time-averaged power can be
written as:
\begin{align}
\left\langle Q\right\rangle = \frac{\omega\eps_0}{4} \int i
\vec{E}_\omega^\ast \cdot  \left(   \db{\eps}^\dagger - \db{\eps}
\right)\cdot \vec{E}_\omega~dV,
\end{align}
and is nonzero only if the permittivity tensor is non-Hermitian
\cite{landau_lifshitz_electrodynamics,agranovich_crystal_1984}.

For the particular case of a material described by the permittivity
tensor \eqref{E:asymmetric_permittivity} (with real-valued
permittivity) one gets
\begin{align}  \label{E:time_averaged_power_transferred_macro}
\left\langle Q\right\rangle  &= \frac{ \omega\eps_0\eps_{xz}}{2} \int      \mathrm{Im} \left\{   E_{\omega,x}^\ast  E_{\omega,z}  \right\}   ~dV
\end{align}
%
% % Sylvain: please check if the above formula needs to be updated. Thank you.
%Professor: No, this one is okay
%
As seen the net power exchanged between the wave and the material is
proportional to $\eps_{xz}$ and its sign depends on the relative
phase between the  components $x$ and $z$ of the electric field.

\section{Exchange of energy between the wave and the material: microscopic approach} \label{sec:microscopic_origin_power_transferred}

Let us now consider a nonlinear material with a symmetric electric 
susceptibility ($\db{\chi}= \db{\chi}^T$) that depends linearly on
the electric field $\db{\chi}=\db{\chi}(\vec{E})$. For such a
material the polarization vector in the time-domain is simply
$\vec{P}(t)=\eps_0\db{\chi} \cdot \vec{E}(t)$
\cite{boyd_nonlinear_2008}. From \eqref{E:power_transferred} it
follows that the instantaneous power transferred from the wave to
the material is
\begin{align}  \label{E:original_power_transferred}
Q= \eps_0 \int \frac{\partial}{\partial t} \left[ \db{\chi} \cdot \vec{E} \right]\cdot \vec{E}~dV.
\end{align}
Since by hypothesis $\db{\chi}$ is a symmetric tensor, the
transferred power can be written as
\begin{align} \label{E:power_transferred_2}
Q= \frac{\eps_0}{2} \frac{d}{d t}\left( \int  \left[ \vec{E} \cdot \db{\chi}  \cdot \vec{E}\right] ~dV \right) + \frac{\eps_0}{2} \int   \vec{E}  \cdot \frac{\partial \db{\chi} }{\partial t} \cdot \vec{E}~dV.
\end{align}
Here the susceptibility depends implicitly on time because of the
nonlinearity:
\begin{align}  \label{E:susceptibility_derivative}
\frac{\partial \db{\chi}}{\partial t}=\sum_{k=x,y,z}\frac{\partial
E_k}{\partial t} \frac{\partial \db{\chi}}{\partial  E_k}
\end{align}
In stationary (time-harmonic) regime, the time average of the first
term in Eq. \ref{E:power_transferred_2} is zero. Then, we can
write:
\begin{align}
\left\langle Q \right\rangle=    \frac{\eps_0}{2} \left\langle  \int   \vec{E}  \cdot \frac{\partial \db{\chi} }{\partial t} \cdot \vec{E}~dV\right\rangle.
\end{align}
Next we decompose the electric field as $\vec{E}=\vec{E}_0
+\vec{E}_\text{AC}$ with $\vec{E}_0$ a static field and
$\vec{E}_\text{AC}$ a time-harmonic field oscillating at the
frequency $\omega$. The time average operation suppresses all the
oscillating terms, so that after simplifications one obtains:
\begin{align}
\left\langle Q \right\rangle & \approx   \eps_0 \left\langle  \int     \vec{E}_\text{AC}  \cdot \frac{\partial  }{\partial t} \left[ \db{\chi} \cdot \vec{E}_0  \right]  ~dV\right\rangle \\
& \approx     \left\langle  \int     \vec{E}_\text{AC}  \cdot \vec{j}^\text{AC-DC} ~dV\right\rangle
\end{align}
Comparing this result with equations \eqref{E:power_transferred} and
\eqref{E:original_power_transferred}, we see that
$\vec{j}^\text{AC-DC}=\eps_0\partial_t \left[ \db{\chi} \cdot
\vec{E}_0  \right]$ defines a dynamic polarization current that is
at the origin of the power exchanged between the time-harmonic field
and the material. This polarization current is induced by the
nonlinearity and by the static electric field.

For the MOSFET metamaterial, the susceptibility is given by Eq. (1) of the main text, so that only the $
\chi_{xx}$ component depends on the electric field. From Eq.
\eqref{E:susceptibility_derivative} a $z$-oscillating field induces
a polarization current along the $x$-direction:
\begin{align}
\vec{j}^\text{AC-DC}= \eps_0 \frac{\partial \chi_{xx}}{\partial  E_z} \frac{\partial E_z}{\partial t} E_{0x} \hat{\vec{x}}
\end{align}
where $E_{0x} =  \vec{E}_0\cdot\hat{\vec{x}}$. Hence, writing
$\vec{E}_\text{AC}=\frac{1}{2}\left(\vec{E}_\omega \e{-i\omega t} +
\vec{E}_\omega^\ast \e{i\omega t}\right)$ the net power exchanged
between the wave and the material becomes:
%Assuming that $\chi_{xx}$ depends only linearly on $\vec{E}$ and that the dynamic field is given by $\vec{E}_\text{AC}=\frac{1}{2}\left(\vec{E}_\omega \e{-i\omega t} + \vec{E}_\omega^\ast \e{i\omega t}\right)$, we obtain that the power transferred is
%
\begin{align}
\left\langle Q \right\rangle &\approx   \frac{\eps_0\omega}{2}  \frac{\partial \chi_{xx}}{\partial  E_z}  E_{0x}     \int  \mathrm{Im} \left\{    E_{\omega,x}^\ast   E_{\omega,z}      \right\}~dV.
\end{align}
%
% Sylvain: please check the above formula (I did not change it)
%Professor: I think it is okay
%
As shown in the main text the linearization of the electromagnetic
response leads to $\eps_{xz}=\frac{\partial \chi_{xx}}{\partial E_z}
E_{0x}$. The above formula coincides with Eq.
\eqref{E:time_averaged_power_transferred_macro}, showing that the
microscopic and macroscopic approaches yield precisely the same
result.

\section*{References}
% \bibliographystyle{apsrev4-1}
% \bibliography{Biblio_CLEAN}